\documentclass[aps,prl,twocolumn,showpacs]{revtex4}
\usepackage{color}
\usepackage{graphicx}
\usepackage{ulem}

\newcommand{\nrl}[1]{{#1}_{\mathcal{Z}}}% single noise realisation
\newcommand{\ihq}{\frac{{{i}}}{{\hbar}}}% i over hbar
\newcommand{\EW}[1]{\left\langle #1 \right\rangle}% qm expectation value
\newcommand{\sEW}[1]{{\mathbf{E}}\left[ #1 \right]}% stochastic average

\begin{document}

\title{Heat flux and information backflow in cold condensed matter systems}
\date{\today}
\author{R. Schmidt$^{1,2}$, S. Maniscalco$^{2}$ and T. Ala-Nissil{a}$^{1,3}$}
\affiliation{$^1$Center for Quantum Engineering and COMP Center of Excellence, Department of Applied Physics, Aalto University School of Science, P.O. Box 11000,
FIN-00076 Aalto, Finland\\
$^2$Turku Centre for Quantum Physics, Department of Physics and Astronomy, University of Turku, FIN-20014 Turku, Finland\\
$^3$Department of Physics, Brown University, Providence RI 02912-1843, U.S.A.}
%\ead{Rebecca.Schmidt@aalto.fi}

\begin{abstract}
We examine non-Markovian effects in an open quantum system from the point of view of information flow. 
To this end, we consider the spin-boson model with a cold reservoir, accounting for the exact time-dependent correlations between the system and the bath to study the exchange of information and heat.
We use an information theoretic measure of the relevant memory effects and demonstrate that the
information backflow from the reservoir to the system does not necessarily correlate with the backflow of heat. 
We also examine the influence of temperature and coupling strength on the loss and gain of information between
the system and the bath. Finally,  we discuss how additional driving changes the backflow of information, giving rise to potential 
applications in reservoir engineering.
\end{abstract}

\maketitle

\paragraph{Introduction.}
In all practical applications quantum systems are open and there is coupling to an external environment or a reservoir, a heat bath. In modeling open quantum systems the environment considered is usually memoryless i.e. Markovian, and therefore detrimental for any quantum coherences. While it is well known for many condensed matter settings that the Markovian approximation does not hold, this fact  has mainly been  considered as a nuisance, giving rise to additional mathematical complexity. Recently, however, a number of results have appeared in the literature indicating that non-Markovian dynamics and, more precisely, the occurrence of information backflow in the system, may be seen as a resource for certain specific information tasks \citep{Vasile2011,Huelga2012,Laine2014,Bylicka2014,Benedetti2014}. The possibility of using information backflow in combination of reservoir engineering techniques motivates the use of non-Markovianity measures or quantifiers, as those defined in Refs. \citep{BLP09,Huelga2012,Chrus2014, Liu2011}.
The systems studied in this context typically involve a structured environment, resulting in time-dependent decay rates in the effective master equation. To the contrary, heat baths considered as environments in standard condensed matter settings do not typically have structured spectral densities. The spin-boson model [17,18] in a cold environment, considered in this Letter, is a paradigmatic example of this situation. Despite the undoubted importance of this model in condensed matter physics,  the question of whether or not this system exhibits non-Markovianity in the information-theoretical sense has not yet been answered. This is one of the main goals of the present Letter.

%Figure 1: Theme
\begin{figure}
  \centering
 \includegraphics[width=0.8\linewidth]{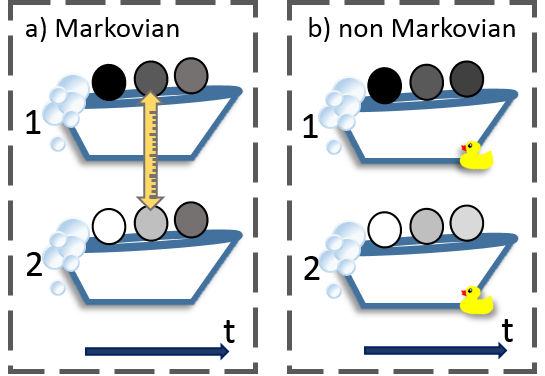}\\
  \caption{(Color online) (a) The distinguishability of two initial preparations (black and white) of an open system decays in time (shades of grey) if the dynamics is Markovian and eventually vanishes in the absence of decoherence free subspaces. 
  (b) For non-Markovian dynamics, the distinguishability can increase again during the propagation due to
  information flow between the two systems.}\label{fig:theme}
\end{figure}

Driven open quantum systems have been in the focus recently also with respect to non-equilibrium quantum thermodynamics \cite{Camp09,Esposito2009,campisi:2011,Hekking13,solinas:2013,Pekola2013,rossnagel:2014,Roncaglia2014,Kutvonen2015}. Here the importance of non-Markovian
effects is highlighted by recent experimental \cite{Serreli2007,Toyabe2010,Roldan2014,Koski2014,Koski2014a,Koski2015} and theoretical \cite{Parrondo2015,Kutvonen2016,Kutvonen2016a,Strasberg2013} works on realising the ubiquitous Maxwell's Demon in strongly coupled single electron and qubit devices. Also, a theoretical framework has been developed based on the concept of a non-equilibrium  subsystem, where some of the degrees of freedom of the reservoir are driven out of equilibrium by the system-bath correlations, leading to extra entropy production terms \cite{Kutvonen2015}. However, entropy alone does not constitute a good measure of information exchange \citep{Bylicka2015}. Therefore, proper information theoretic tools and their relation to heat exchange should be examined.

In the present Letter we consider a simple but realistic model, the coupled spin-boson quantum system, where the
bosonic heat bath has a well-defined spectral density and its properties can be adjusted by changing its temperature. We use the stochastic Liouville-von Neumann scheme to study the dynamics
of the spin-boson system  at low temperatures, accounting for the exact system-bath
correlations giving rise to non-Markovian effects.  We employ the Breuer-Laine-Piilo (BLP) measure \cite{BLP09} to quantify non-Markovianity. Our results demonstrate that
non-Markovian effects arise in the driven spin-boson model at low temperatures, and are not necessarily correlated with
the exchange of heat between the system and the bath. We also discuss how external driving could be used as a resource 
in quantum environmental engineering.

%Theory
\paragraph{The Model.}
We consider here a simple but realistic model where a two level system (TLS) 
$H_{S}=- (\hbar\omega/{2})\sigma_{x}$ is coupled to a bosonic heat bath \cite{weiss2008,breuer:2007}, 
where $\omega$ is the system frequency and $\sigma_x$ is a spin Pauli matrix. 
The full Hamiltonian is given by $H=H_{S}+H_{I}+H_{R}$, 
with the reservoir and the interaction Hamiltonians $H_R=\sum_k \hbar\omega_k b_k^\dagger b_k$ and 
$H_{I}=\sigma_z \sum_k c_k (b_k^\dagger+b_k)$, respectively. The reservoir is modelled by a large number of 
quantum harmonic oscillators with frequencies $\omega_k$, as well as the 
annihilation $b_k$ and creation $b_k^\dagger$ operators. The impact of the reservoir on the system depends only 
on its thermal energy $k_{\rm B}T\equiv 1/\beta$ and spectral density function $J(\omega)$. 
Here, we consider an Ohmic spectral density with a large algebraic cut-off $\omega_c$,  $J(\omega)=\hbar \gamma \omega /(1 + (\omega /\omega_c)^2)^2$, where $\gamma$ is a dimensionless coupling constant \cite{weiss2008}\footnote{Note: The definition of $\gamma$ \cite{weiss2008} deviates from \cite{breuer:2007} by a factor of $1/2$.}.
Unlike in many other quantum information theoretic studies, the cut-off frequency $\omega_c$ here is chosen to be large enough
($\omega_c = 10 \omega)$ such that it is not the source of non-Markovian effects in our system. Nevertheless, for low temperatures and large coupling strength (typical of, e.g., superconducting devices), the Born-Markov approximation is not applicable, as the dynamics of the open system becomes non-local in time.  An exact equation for the reduced density matrix of the system $\rho(t)$ can be derived from the path integral formalism \cite{weiss2008} and it is known as the stochastic Liouville-von Neumann equation (SLN) \cite{Stock2002,Stock2004}:   
\begin{equation}\label{sln}
\nrl{\dot{\rho}}(t)= -\ihq [H_{S}(t),\nrl{\rho}]+\ihq \xi(t) [\sigma_{z},\nrl{\rho}]+\frac{i}{2}\nu(t)\{\sigma_{z},\nrl{\rho}\}\, .
\end{equation}
This equation holds for a single noise realisation $\mathcal{Z}\equiv\{\xi,\nu\}$. The correlation functions of the two complex-valued noise forcing terms $\xi(t)$ and $\nu(t)$ reproduce the complex-valued and temporally non-local force-force autocorrelation function of the bath. Therefore, the memory effects of the dynamics are embedded into the noise correlations, while Eq. (\ref{sln}) is local in time.
The physical, reduced density operator $\rho(t)$ is obtained as an expectation value 
over a large number of noise realisations, i.e.,
\begin{equation}\label{avrho}
\rho(t) = \sEW {\nrl{\rho}(t)},
\end{equation}
In the following, we consider resonant, periodic driving of the system, which only changes the system 
Hamiltonian $H_{S} \rightarrow H_{S}(t)$:
\begin{equation}\label{driving}
H_{S}(t)=H_0+H_D(t)=- \frac{\hbar\omega}{2}\sigma_{x}+\lambda_0\, \sin(\omega \,t)\,  \sigma_z\, ,
 \end{equation}
as the SLN treats the system-bath interaction exactly. The driving couples to the same system degree of freedom as the bath and therefore acts as an additional contribution to the reservoir. Therefore, the driving is in essence reservoir engineering.

The non-equilibrium thermodynamics of this system has recently been studied in Ref. \cite{RS2015}.  The heat flux between the 
system and environment is given by
\begin{equation}\label{heatflux}
j_{Q}(t)=-\omega\ \sEW{\xi(t)\EW{\sigma_y (t)}}\, ,
\end{equation}
where the $\EW{\cdot}$ denote the quantum mechanical average.
The derivation of Eq. (\ref{heatflux}) is based on the definition of work via the power operator, 
as introduced in Ref. \cite{solinas:2013}. In the Heisenberg picture, the first law of thermodynamics in the 
Hilbert space of the composite system gives the heat flow (for details see Ref. \citep{RS2015}).    
Throughout this paper, we use natural units where $\omega =1$, $\hbar=1$ and $k_B =1$. 

\paragraph{Measure of Non-Markovianity.} 
There are several different approaches to quantify the information backflow in non-Markovian dynamics  \citep{BLP09,Huelga2012,Bylicka2014,Chrus2014, Liu2011}.  
The BLP measure of Ref. \cite{BLP09} monitors the dynamics of distinguishability between two initial preparations. 
In the Markovian case, the open system dynamics monotonically decreases the distinguishability which eventually vanishes for dynamics with a unique steady state (cf. Fig. \ref{fig:theme}). 
To quantify the distinguishability, the BLP measure employs the trace distance $D$ as
\begin{equation}
D(\rho_{1},\rho_{2})=\frac{1}{2}{\rm{Tr}}|\rho_{1}-\rho_{2}|
\end{equation}
The information flow $\Delta$ is defined as the change of the trace distance in time, and is given by
\begin{equation}\label{Delta}
\Delta(t,\rho_{1} (0),\rho_{2}(0))=\frac{{\rm{d}}}{{\rm{d}}t}D(\rho_{1},\rho_{2})\, .
\end{equation}
Only time intervals where the trace distance increases, contribute to the BLP measure:
\begin{equation} \label{BLP}
\mathcal{N}(\Phi)=\max_{\rho_{1},\rho_{2}}\int_{\Delta>0}dt \Delta(t,\rho_{1},\rho_{2}).
\end{equation} 
This measure also includes an optimization over all possible input states. It is known that for a TLS, 
the pair of states that maximises the BLP measure is pure and located on the opposite sides of the 
Bloch sphere \cite{Wissmann12}. The optimal pair depends on the propagation time. 
In the following we concentrate on $\Delta$ (cf. Eq. (\ref{Delta})) instead of the full BLP measure of Eq. (\ref{BLP}). 
While avoiding the numerically demanding optimization, this also provides more insight into the 
dynamics of information backflow and how it can be controlled by means of an external drive.  
Given that the orientation of the system is not completely random, as the bath couples to $\sigma_{z}$ and 
therefore distinguishes this basis, we will examine the behaviour of the trace distance of the plus and minus eigenstates of the Pauli matrices. In other words, for each Pauli matrix we choose its pair of eigenvectors as the initial states, and then compute their respective dynamics and the corresponding information flow $\Delta$. 

%Results
\paragraph{Information Backflow.}
%Figure 2: backflow of information
\begin{figure}
  \centering
 \includegraphics[width=\linewidth]{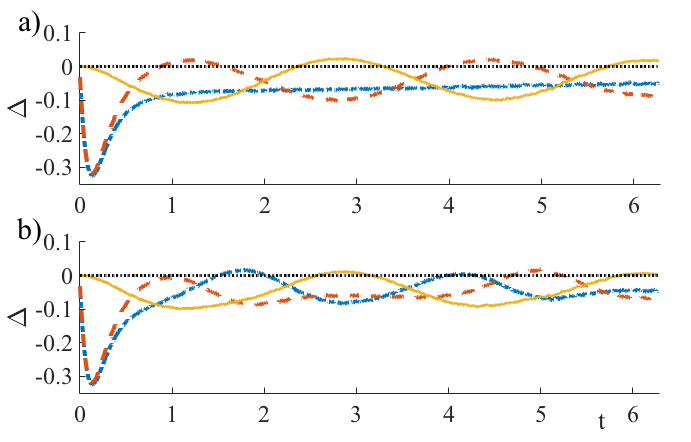}\\
  \caption{(Color online) Information flow as measured by the quantity $\Delta$ (cf. Eq. (6)) during propagation time. 
  Positive values of $\Delta$ (data above the dashed horizontal black line) correspond to information backflow from the bath to the system. The different curves correspond to different initial preparations of the system, where the two states are the eigenstates of each spin
  operator (blue dash-dotted line for $\sigma_{x}$, 
  red dashed line for $\sigma_{y}$, and yellow solid line for $\sigma_{z}$). All data are 
  for $\gamma =0.05$ and $\beta=5$. The upper panel
  (a) shows the dynamics of the system without an external drive, while in the lower panel (b) the system is driven periodically with a resonant drive of amplitude $\lambda_{0}=1$.}\label{fig:backflow}
\end{figure}

In Fig. \ref{fig:backflow} we plot the time evolution of the information flow for different initial preparations of the driven and undriven cases. The upper panel displays the case without an external drive. It can be seen that except for the case where the initial preparation corresponds to the eigenstates of the bare Hamiltonian, $\sigma_x$, there are time windows during the relaxation where there are positive values of $\Delta$, corresponding to backflow of information from the bath to the system. This figure shows clearly that the dynamics of the system is non-Markovian in the sense of the BLP measure. 
The lower panel of  Fig. \ref{fig:backflow} shows data for the same parameter set, but with periodic, resonant driving, with driving amplitude $\lambda_0=1$. The drive has a clear influence on the memory effects. It reduces the backflow of information seen in the upper panel for eigenstates of $\sigma_y$ and $\sigma_z$, but now there is also information backflow for the initial preparation in eigenstates of $\sigma_x$.

\paragraph{Heat flow.}
%Figure 3: Backflow and heatflux
\begin{figure}
  \centering
 \includegraphics[width=\linewidth]{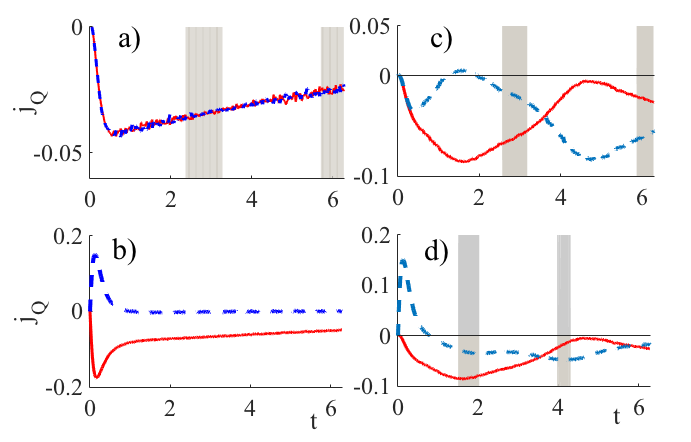}\\
  \caption{(Color online) Heat flux of Eq. (4) for different initial preparations of the system, where the upper panels (a) and (c) are
  for the two eigenstates of $\sigma_{z}$, and (b) and (d) are for the two eigenstates of $\sigma_{x}$ (red solid lines are for the $+$ 
  spin eigenstates and blue dashed lines for the $-$ states). All data are for $\gamma =0.05$ and $\beta=5$. Panels (a) and (b)
  are for the undriven case, and panels (c) and (d) for the same drive as in Fig. 2. 
  Negative values of $j_Q(t)$ correspond to a heat flux from the system into the environment.  Grey shaded areas indicate
  time windows of information backflow for the corresponding trace distance, as shown in Fig. \ref{fig:backflow}.}\label{fig:heat}
\end{figure}

Next we consider the heat flux between the system and the reservoir, as defined in Eq. (\ref{heatflux}). In this setting,  the heat flux does not neccessarily flow unidirectionally from the system to the environment but heat can also return into the system (see, e.g., \cite{RS2015}). 
Most importantly, the results in Fig. 3 show that there does not have to be any correlation between heat and information backflow for either the driven or undriven cases. Even for initial pairs in the driven case, where both information and heat flow back during the propagation, this does not happen simultaneously. While the dynamics is non-Markovian (i.e. there are initial pairs which show information backflow) and there is heat back flow into the system for certain (other) initial preparations as in \cite{Guarnieri2016}, we do \emph{not} find any correlations between heat flux and information backflow for the same pair of initial states.

\paragraph{Non-Markovianity as a Resource.} 
%
%Figure 4: Loss and gain
\begin{figure}
  \centering
 \includegraphics[width=\linewidth]{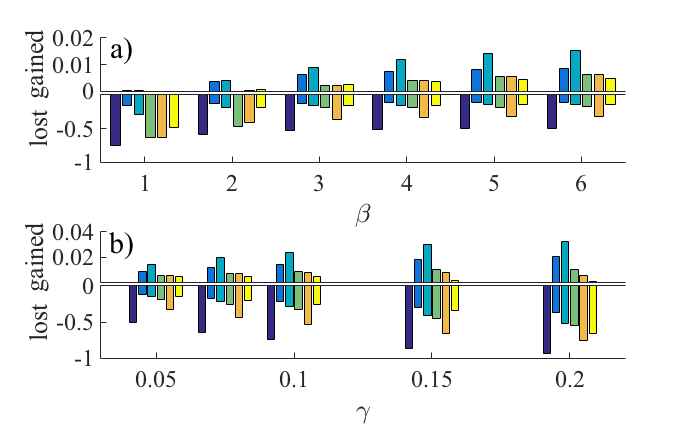}\\
  \caption{(Color online) Information lost and gained during the propagation for different values of  the inverse temperature 
  $\beta$ (a) and the coupling strength $\gamma$ (b), with and without driving. The negative values of $I_{\Delta}(t)$ correspond
  to the information lost before the first backflow event occurs, while the positive bars quantify the amount of information
  flowing back (if any) for the first time. The colors in the bars correspond to trace distances calculated between the two eigenstates
   of undriven states $\sigma_{x}$ (dark blue), $\sigma_{y}$ (blue), and $\sigma_{z}$ (light blue), and driven eigenstates $\sigma_{x}$ (green), $\sigma_{y}$ (orange), and $\sigma_{z}$ (yellow). Note the different scales on the vertical axes for loss and gain.}\label{fig:gain}
\end{figure}

The information backflow shown in Fig. \ref{fig:backflow} (and for a wider range of parameters in Fig. \ref{fig:gain}) is quantitatively
 small, in particular in relation to the information lost before the information flow changes direction. To use non-Markovianity as a resource, a deeper understanding of how loss and gain of distinguishability are influenced by the parameters of the system and the possibilities to change this with additional driving is required. To quantify the information exchange, we define the quantity
 \begin{equation}
 I_{\Delta}(t) = D(\rho_{1}(t_{}),\rho_{2}(t_{}))-D(\rho_{1}(0),\rho_{2}(0)),
 \end{equation}
 which measures the loss of information (difference between the trace distances) after time $t$ from the beginning of the
 dynamics. We note that in the present case $D(\rho_{1}(0),\rho_{2}(0))=1$ as we always start from orthogonal eigenstates.
 
 Figure \ref{fig:gain} shows the information loss that occurred before the distinguishability increased 
 for the first time for a range of temperatures and coupling strengths. 
 If there is no information backflow (as for all undriven cases with initial preparations of 
 eigenstates for $\sigma_x$), the lower part of each subfigure, respectively of Fig. \ref{fig:gain} shows the information loss during the full propagation time ($2\pi$). With decreasing bath temperature, i.e., when the dynamics becomes more non-Markovian, the overall information loss decreases 
 and the subsequent partial information regain increases. For increasing coupling constant the picture is more complicated, 
 since both the backflow of information and the loss increase. Adding an external drive alters the general picture. While the information backflow for the $\sigma_z$ initial eigenstates is suppressed, an increase in the distinguishability of the $\sigma_x$ initial eigenstates occurs. There might be experimental situations where storing the information in the eigenstates of the bare system Hamiltonian is more favourable. Our results show that tailored driving offers the possibility to enhance non-Markovianity, and hence the backflow of information, in the desired direction. The first instance where the information flow is reversed (not shown), depends only very weakly on $\beta$ and $\gamma$, but changes considerably when driving is present as can for example be seen for the set of parameters in Fig. \ref{fig:backflow}. 

\paragraph{Summary.}
We have studied for the first time the information theoretic concept of non-Markovianity in a paradigmatic condensed matter system, i.e. the exact spin-boson model, finding that backflow of information does occur in the system. Our results show that there is generally no connection between information exchange and heatflow between the system and the bath either for the undriven or driven cases.  We have also examined the influence of
temperature and coupling strength on information loss and regain in the model. Finally, we have provided insight in how the information backflow can be influenced by driving. Our results thus pave the way to a follow-up investigation, where we plan to use optimal control techniques to tackle this question in its full generality.

\acknowledgements{We thank Jukka Pekkola for fruitful discussions. We gratefully acknowledge financial support by the COST Action MP1209, the Center of Quantum Engineering at Aalto University School of Science, the EU Collaborative project QuProCS (Grant Agreement 641277), the Academy of Finland through its Centres of Excellence Programme (2015-2017) under project numbers 284621 and 287750, and the Magnus Ehrnrooth Foundation.}

%%%%%%%%%%%%%%%%%%%%%%%%%%%%%%%%%%%

\bibliographystyle{apsrev4-1}
\bibliography{Literature_NoMaTh}
\end{document}